\documentclass[twocolumn,twoside]{article}

\usepackage{graphicx}
\usepackage{hyperref}

\begin{document}

\thispagestyle{plain} \markboth{\rm ON PARTICLES COLLISIONS NEAR
ROTATING BLACK HOLES }{\rm GRIB, PAVLOV}

\twocolumn[
\begin{center}
{\LARGE \bf On Particles Collisions Near Rotating Black Holes}
\vspace{12pt}

{\Large {A. A. Grib}${\,}^{1,\, 2\,{\scriptstyle *}}$, \ \
{Yu. V. Pavlov}${\,}^{1,\,3,\dag}$} \vspace{12pt}

{\it ${}^1$A. Friedmann Laboratory for Theoretical Physics,\\
Griboedov kanal 30/32, St.\,Petersburg 191023, Russia}
\vspace{4pt}

{\it ${}^2$Russian State Pedagogical University (The Herzen University),\\
Moyka-river emb. 48, St. Petersburg 191186, Russia}
\vspace{4pt}

{\it ${}^3$Institute of Mechanical Engineering, Russian Acad. Sci.,\\
Bol'shoy pr. 61, St. Petersburg 199178, Russia}
\end{center}
\vspace{5pt}
 {\bf Abstract.}
    Scattering of particles with different masses and energy in
the gravitational field of rotating black holes is considered
as outside as inside the black hole.
    Expressions for scattering energy of particles in the centre of mass
system are obtained.
    It is shown that scattering energy of particles in the centre of mass
system can obtain very large values not only for extremal black holes
but also for nonextremal ones if one takes into account multiple scattering.
    Numerical estimates for the time needed for the particle to get
ultrarelativistic energy are given.

\vspace{11pt}
%% 04.70.-s Physics of black holes
%% 04.70.Bw Classical black holes
%% 97.60.Lf Black holes (in astronomy)
\vspace{25pt}
]

%%%%******************************************************************
{\centering  \section{\uppercase{\rm Introduction}}}

\footnotetext[1]{E-mail: \, andrei\_grib@mail.ru}
\footnotetext[2]{E-mail: \, yuri.pavlov@mail.ru}

    In our publications~\cite{GribPavlov2010} we came to the conclusion that
one can get very large energies in the centre of mass frame first mentioned
in~\cite{BanadosSilkWest09} for two particle collision of particles with
equal masses close to the horizon of the rotating black hole
(Active Galactic nucleus) if one considers multiple scattering.
    The effect of getting very large energy depends on the value of the
angular momentum of one of the particles.
    The problem of the energy of collision of particles in vicinity of
black holes of different types now is intensively studied by different
authors~\cite{WeiLiuGuoFu10,Zaslavskii10,MaoLiJiaRen10}.
    Here we obtain similar formulas for particles of different masses
in the field of Kerr's black hole.
    To get very large energy one must have large time of rotating of the particle
around the black hole coming closer and closer to the horizon.
    We give some quantitative estimates for the time needed for a particle
to obtain ultrarelativistic energy outside the horizon.
    Then we investigate the case of scattering inside the horizon.
    The limiting formulas are obtained and it is shown
that the collisions with infinite energy can not be realized
even in the singularity.

    The system of units $G=c=1$ is used in the paper.

%\vspace{17pt}
%%%%%%%%%%%%%%%% Section "Energy of Collision in Black Holes" %%%%%%%%
{\centering \section{\uppercase{\rm
The energy of collisions in the field of Kerr's black hole}}
\label{2secBHColl}
}

    Let us consider particles falling on the rotating chargeless black hole.
    The Kerr's metric of the rotating black hole in Boyer--Lindquist
coordinates has the form                    %% R.H.Boyer, R.W.Lindquist
    \begin{eqnarray}
d s^2 = d t^2 -
\frac{2 M r \, ( d t - a \sin^2 \! \theta\, d \varphi )^2}{r^2 + a^2 \cos^2
\! \theta } - (a^2 \cos^2 \! \theta
\nonumber \\
+\, r^2 ) \Bigl( \frac{d r^2}{\Delta} + d \theta^2 \Bigr)
- (r^2 + a^2) \sin^2 \! \theta\, d \varphi^2,
\label{Kerr}
\end{eqnarray}
    where
    \begin{equation} \label{Delta}
\Delta = r^2 - 2 M r + a^2,
\end{equation}
    $M$ is the mass of the black hole, $J=aM$ is angular momentum.
    In the case $a=0$ the metric~(\ref{Kerr}) describes the static chargeless
black hole in Schwarzschild coordinates.
    The event horizon for the Kerr's black hole corresponds to the value
    \begin{equation}
r = r_H \equiv M + \sqrt{M^2 - a^2} \,.
\label{Hor}
\end{equation}
    The Cauchy horizon is
    \begin{equation}
r = r_C \equiv M - \sqrt{M^2 - a^2} \,. \label{HorCau}
\end{equation}
%%%%%%%%%%%%%%%%%%%%%%%%%%%%%%%%%%%%%%%%%%%%%%%%%%
    For equatorial ($\theta=\pi/2$) geodesics in Kerr's metric~(\ref{Kerr}) one
obtains (\cite{Chandrasekhar}, \S\,61):
    \begin{equation} \label{geodKerr1}
\frac{d t}{d \tau} = \frac{1}{\Delta} \left[ \left(
r^2 + a^2 + \frac{2 M a^2}{r} \right) \varepsilon - \frac{2 M a}{r} L \right],
\end{equation}
    \begin{equation}
\frac{d \varphi}{d \tau} = \frac{1}{\Delta} \left[ \frac{2 M a}{r}\,
\varepsilon + \left( 1 - \frac{2 M}{r} \right)\! L \right],
\label{geodKerr2}
\end{equation}
    \begin{equation} \label{geodKerr3}
\left( \frac{d r}{d \tau} \right)^2 = \varepsilon^2 +
\frac{2 M}{r^3} \, (a \varepsilon - L)^2 +
\frac{a^2 \varepsilon^2 - L^2}{r^2} - \frac{\Delta}{r^2}\, \delta_1 ,
\end{equation}
    where
    $\delta_1 = 1 $ for timelike geodesics
($\delta_1 = 0 $ for isotropic geodesics),
$\tau$ is the proper time of the moving particle,
$\varepsilon={\rm const} $ is the specific energy:
the particle with rest mass~$m$ has the energy $\varepsilon m $ in the
gravitational field~(\ref{Kerr});
$ L m = {\rm const} $ is the angular momentum of the particle relative
to the axis orthogonal to the plane of movement.

%%%%%%%%%%%%%%%%%%%%%%%%%%%%%%%%%%%%%%%%%%%%%%%%%%
    Let us find the energy $E_{\rm c.m.}$ in the centre of mass system
of two colliding particles with rest masses~$m_1$ and~$m_2$
in arbitrary gravitational field.
    It can be obtained from
    \begin{equation} \label{SCM}
\left( E_{\rm c.m.}, 0\,,0\,,0\, \right) =
m_1 u^i_{(1)} + m_2 u^i_{(2)}\,,
\end{equation}
    where $u^i=dx^i/ds$.
    Taking the squared~(\ref{SCM}) and due to $u^i u_i=1$ one obtains
    \begin{equation} \label{SCM2}
\frac{E_{\rm c.m.}^{\,2}}{2 m_1 m_2} = \frac{m_1}{2 m_2} + \frac{m_2}{2 m_1}
+ u_{(1)}^i u_{(2) i} \,.
\end{equation}
    The scalar product does not depend on the choice of the coordinate frame
so~(\ref{SCM2}) is valid in an arbitrary coordinate system and for arbitrary
gravitational field.
%%%%%%%%%%%%%%%%%%%%%%%%%%%%%%%%%%%%%%%%%%%%%%%%%%

    We denote~$x=r/M$, \ $ A=a/M$, \ $ l_n=L_n/M$,
\ $ \Delta_x = x^2 - 2 x + A^2 $     and
    \begin{equation} \label{DenKBHxhc}
x_H = 1 + \sqrt{1 - A^2}, \ \ \
x_C = 1 - \sqrt{1 - A^2}.
\end{equation}
    For the energy in the centre
of mass frame of two colliding particles with specific energies
$\varepsilon_1$,  $ \varepsilon_2 $ and
angular momenta $L_1, \, L_2$,
which are moving in Kerr's metric one obtains
using~(\ref{Kerr}), (\ref{geodKerr1})--(\ref{geodKerr3}), (\ref{SCM2}):
    \begin{eqnarray}
\frac{E_{\rm c.m.}^{\,2}}{2\, m_1 m_2} = \frac{m_1^2 + m_2^2}{2 m_1 m_2} -
\varepsilon_1 \varepsilon_2 +
\frac{1}{x \Delta_x} \Biggl[ l_1 l_2 (2\!-\!x)
\nonumber \\
+ 2 \varepsilon_1 \varepsilon_2 \biggl( x^2 (x\!-\!1) + A^2 (x\!+\!1) -
A \Bigl( \frac{ l_1}{\varepsilon_1} + \frac{ l_2}{\varepsilon_2} \Bigr)
\!\biggr)
\nonumber \\
-\, \sqrt{ 2 \varepsilon_1^2 x^2 + 2 (l_1 - \varepsilon_1  A )^2 \!- l_1^2 x
+ (\varepsilon_1^2 \!- 1 ) x \Delta_x } \ \
\nonumber  \\
\times \sqrt{ 2 \varepsilon_2^2 x^2 + 2 (l_2 - \varepsilon_2  A )^2 \!- l_2^2 x
+ (\varepsilon_2^2 \!- 1 ) x \Delta_x }\, \Biggr].
\label{KerrL1L2}
\end{eqnarray}
    It corresponds to results in~\cite{BanadosSilkWest09}
and~\cite{GribPavlov2010c} for the case $ \varepsilon_1 = \varepsilon_2$.
    Collisions of particles of equal masses with different specific
energies were considered in~\cite{HaradaKimura10}.

    Writing the right hand side~(\ref{KerrL1L2}) as
$ f(x) + (m_1^2 + m_2^2)/ 2 m_1 m_2 $, one obtains
    \begin{equation} \label{Kerrm1m2max}
\frac{E_{\rm c.m.}(r)}{m_1 + m_2} =
\sqrt{ 1 + (f(x) -1) \frac{2 m_1 m_2}{(m_1 + m_2)^2}}.
\end{equation}
    This relation has maximal value for given $r$, specific particle
energies $ \varepsilon_1, \,  \varepsilon_2 $ and specific angular momenta
$ l_1, \, l_2 $, if the particle masses are equal: $m_1 =m_2$.

    To find the limit $r \to r_H$ for the black hole with a given angular
momentum~$A$ one must take in~(\ref{KerrL1L2}) $x = x_H + \alpha$
with $\alpha \to 0 $ and do calculations up to the order~$\alpha^2$.
    Taking into account $ A^2 = x_H x_C$, $x_H + x_C=2$, after resolution
of uncertainties in the limit $\alpha \to 0 $ one obtains
    \begin{eqnarray}
\frac{E_{\rm c.m.}^{\,2}(r \to r_H)}{2\, m_1 m_2} =
\frac{m_1^2 + m_2^2}{2 m_1 m_2} - \frac{l_{1H} l_{2H}}{4}
\nonumber \\
+\, \frac{1}{8} \left[ (l_{1H}^2 +4) \frac{l_{2H} - l_2}{l_{1H} - l_1} +
(l_{2H}^2 +4) \frac{l_{1H} - l_1}{l_{2H} - l_2}
\right],
\label{KerrLimA}
\end{eqnarray}
    where
    \begin{equation} \label{KerrlH}
l_{nH} = \frac{2 \varepsilon_n x_H}{A} =  \frac{2 \varepsilon_n}{A}
\left( 1 + \sqrt{1-A^2} \, \right).
\end{equation}
is the limiting value of the angular momentum of the particle with
specific energy $ \varepsilon_n $ close to the horizon of the black hole.
    It can be obtained from the condition
of positive derivative in~(\ref{geodKerr1}) $dt /d \tau > 0$,
i.e. going ``forward'' in time:
    \begin{equation} \label{KerrLehxa}
l_n < l_{nH} \left( 1 + \frac{x_H +1}{2}\, \alpha \right) + o(\alpha), \ \
x =x_H+ \alpha .
\end{equation}
    So close to the horizon one has the condition
$l_n \le l_{nH} $.

    Note that for $l=l_H - \beta$ from~(\ref{geodKerr3}) one gets
    \begin{equation} \label{geodlHbeta}
\left( \frac{d r}{d \tau} \right)^{\!2}\biggr|_{r=r_H} =
\frac{\beta^2 x_C}{x_H^3} > 0 \,.
\end{equation}
    So there exists some region close to the horizon where one has particles
moving with angular momentum arbitrary close to the limiting value~$l=l_H$.

    In another form~(\ref{KerrLimA}) is
    \begin{eqnarray}
\frac{E_{\rm c.m.}(r \to r_H)}{m_1 + m_2} =
\left[ 1+ \frac{m_1 m_2}{(m_1+m_2)^2} \right. \hspace{22pt}
\nonumber \\
\times \left. \frac{ (l_{1H} l_2 \!-l_{2H} l_1)^2 +
4 (l_{1H} \!-\! l_{2H} + l_2 \!-\! l_1)^2}
{4 (l_{1H}-l_1) (l_{2H} -l_2)} \right]^{1/2} \!\!.
\label{KerrLime1e2e2f}
\end{eqnarray}
    In special case $ \varepsilon_1 = \varepsilon_2 $
(for example for nonrelativistic on infinity
particles $ \varepsilon_1 = \varepsilon_2 =1 $)
formula~(\ref{KerrLimA}) can be written as
    \begin{eqnarray}
\frac{E_{\rm c.m.}(r \to r_H)}{m_1 + m_2} = \hspace{33pt}
\nonumber \\
= \sqrt{ 1+ \frac{m_1 m_2}{(m_1+m_2)^2}
\frac{(4+l_H^2)\, (l_1 - l_2)^2}{4 (l_H-l_1) (l_H -l_2)}}.
\label{KerrLime1e2e}
\end{eqnarray}

    For the extremal black hole $A=x_H=1$, $ l_{nH}=2 \varepsilon_n $ and
the expression~(\ref{KerrLimA})
is divergent when the dimensionless angular momentum of
one of the falling into the black hole particles $ l=l_H=2 \varepsilon$.
    The scattering energy in the centre of mass system is increasing without
limit
(for case $\varepsilon=1$ it was established in~\cite{BanadosSilkWest09}).
    For example, if $l_1=l_{1H} $ then  one obtains from Eq.~(\ref{KerrL1L2})
    \begin{equation} \label{KerrLimAExxx}
\frac{E_{\rm c.m.}^{\,2}(x)}{2\, m_1 m_2} \approx
\frac{ (2 \varepsilon_2 - l_2) ( 2 \varepsilon_1
-\sqrt{ 3 \varepsilon_1^{2} - 1 }\, )  }{x-1} , \ \ \ x \to 1 .
\end{equation}
%%%%%%%%%%%%%%%%%%%% end my %%%%%%%%%%%%%%%%%%%%%%%%%%%%%%%%%%%%%%%%%%
    Note that the small value of $r-r_H$ for the radial coordinate of the
point of the collision of particles with high energy in the
centre of mass frame does not mean small distance
because the metrical coefficient $g_{rr} =-r^2/ \Delta \to \infty$.
%%%%%%%%%%%%%%%%%%%% end my %%%%%%%%%%%%%%%%%%%%%%%%%%%%%%%%%%%%%%%%%%

    If $A=1$ and $l=l_H=2 \varepsilon$ then from~(\ref{geodKerr3})
one gets
    \begin{equation} \label{geodlH}
\left( \frac{d r}{d \tau} \right)^{\!2} =
\frac{(x-1)^2}{x^3} \left( 2 \varepsilon^2 +(\varepsilon^2 - 1) x \right).
\end{equation}
    For $\varepsilon \ge 1$ expression~(\ref{geodlH}) is nonnegative and
the particle with such angular momentum falling from the infinity can
achieve the event horizon.

    In Refs.~\cite{GribPavlov2010,GribPavlov2010c}
we are shown that in order to get the unboundedly growing energy one must
have the time interval (as coordinate as proper time) from the beginning
of the falling inside the black
hole to the moment of collision also growing infinitely.
    Give some quantitative estimates.

    In case of the extremal rotating black hole $A=1$
and the limiting angular momentum $l_1= 2 \varepsilon_1 $
from~(\ref{geodKerr1}), (\ref{geodKerr3}) one obtains
    \begin{equation} \label{telH}
\frac{dt}{dx} = - \frac{M \varepsilon_1 \sqrt{x} (x^2 + x + 2)}{(x-1)^2
\sqrt{ \varepsilon_1^2 (x+2) - x } } \,.
\end{equation}
    So the time of movement in the vicinity of events horizon up to
the point of collision with radial coordinate~$x_f \to x_H=1 $
is
    \begin{equation} \label{telH2}
\Delta t \sim \frac{4 M \varepsilon_1}{(x_f -1)
\sqrt{ 3 \varepsilon_1^2 - 1 } } \,.
\end{equation}
    Taking into account~(\ref{KerrLimAExxx}) one obtains time of movement
before collision with a given value of the energy~$E$ in the centre of
mass frame
    \begin{equation} \label{telH3}
\Delta t \sim \frac{E^2}{m_1 m_2} \,
\frac{2 M \varepsilon_1}{ (2 \varepsilon_2 - l_2)
\sqrt{ 3 \varepsilon_1^2 - x }
\,( 2 \varepsilon_1 - \sqrt{ 3 \varepsilon_1^2 - x } \,) } \,.
\end{equation}
    For $\varepsilon_1 = \varepsilon_2 = 1, \, l_2=0$ one gets
    \begin{equation} \label{telH4}
\Delta t \sim \frac{E^2}{m_1 m_2} \,
\frac{M }{ 2 (\sqrt{2} - 1 )} \approx 6 \cdot 10^{-6} \frac{M}{M_\odot}\,
\frac{E^2}{m_1 m_2} \, {\rm s},
\end{equation}
    where $M_\odot$ is the mass of the Sun.

    So to have the collision of two protons with the energy of the order
of the Grand Unification one must wait for the black hole of the star mass
the time $\sim 10^{24}$\,s,
which is large than the age of the Universe $\approx 10^{18}$\,s.
    However for the collision with the energy $10^3$ larger than
 that of the LHC one must wait only~$\approx 10^{8}$\,s.

    Can one get the unlimited high energy of this scattering energy for
the case of nonextremal black hole?

\vspace{17pt}
%%%% Section "The energy of collisions for nonextremal black hole" %%%
{\centering \section{\uppercase{\rm \large
The energy of collisions for nonextremal black hole}}
\label{3secBHnonextr}
}

    For a particle falling on the black hole from infinity one must
have $\varepsilon \ge 1$.
    In this section we consider the case $ \varepsilon = 1$,
when the particles falling into the black hole are
nonrelativistic at infinity.
    Formula~(\ref{geodKerr3}) leads to limitations on the possible values
of the angular momentum of falling particles:
the massive particle free falling in the black hole with dimensionless
angular momentum~$A$ to achieve the horizon of the black hole must have
angular momentum from the interval
    \begin{equation} \label{geodKerr5}
- 2 \left( 1 + \sqrt{1+A} \right) \!=\!l_L \le l
\le l_R \!=\! 2 \left( 1 + \sqrt{1-A} \right)\!.
\end{equation}

    Putting the limiting values of angular momenta $l_L, l_R$
into the formula~(\ref{KerrLime1e2e}) one obtains the maximal values
of the collision energy of particles freely falling from infinity
    \begin{eqnarray} \label{KerrLimAMax}
\frac{E_{\rm c.m.}^{\, \rm max}(r \to r_H)}{m_1+m_2} = \hspace{44pt}
\\
= \sqrt{1 + \frac{2 m_1 m_2}{(m_1 +m_2)^2}
\frac{(2+\sqrt{1+A} +\sqrt{1-A} )^2}{(1+\sqrt{1-A^2}) \sqrt{1-A^2}}}
\,.
\nonumber
\end{eqnarray}
    For $A=1-\epsilon$ with $\epsilon \to 0$ formula~(\ref{KerrLimAMax})
gives:
    \begin{equation} \label{KerrimAE}
E_{\rm c.m.}^{\, \rm max} = 2 \left( 2^{1/4}+2^{-1/4} \right)
\frac{\sqrt{m_1 m_2}}{\epsilon^{1/4}} +O(\epsilon^{1/4}).
%%\approx \frac{m \cdot 4,06}{\epsilon^{1/4} } \,. %% 4.0602070605...
\end{equation}
    So even for values close to the extremal $A=1$ of the rotating black hole
$E_{\rm c.m.}^{\, \rm max}/ \sqrt{ m_1 m_2} $ can be not very large
as mentioned in~\cite{BertiCardosoGPS09} for the case $m_1=m_2$.
    So for $A_{\rm max} =0.998 $ considered as the maximal possible
dimensionless angular momentum of the astrophysical black holes
(see~\cite{Thorne74}), from~(\ref{KerrLimAMax}) one obtains
$ E_{\rm c.m.}^{\, \rm max} /\sqrt{m_1 m_2}\approx 18.97 $.

    Does it mean that in real processes of particle scattering in
the vicinity of the rotating nonextremal black holes the scattering energy
is limited so that no Grand Unification or even Planckean energies can
be obtained?
    Let us show that the answer is no!
    If one takes into account the possibility of multiple scattering so that
the particle falling from infinity on the black hole with some fixed
angular momentum changes its momentum in the result of interaction with
particles in the accreting disc and after this is again scattering close to
the horizon then the scattering energy can be unlimited.

\begin{figure}[ht]
    \includegraphics[width=79mm]{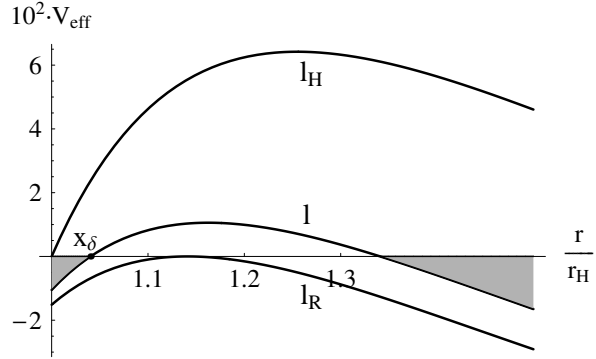}
\caption{ \label{Veff} The effective potential for $A=0.95$ and
$l_R \approx 2.45 $, \ %%2.44721}
$ l=2.5 $, \
$l_H \approx 2.76 $. %%2.76263}.
Allowed zones for \, $ l=2.5 $ are shown by the gray color.
}
\end{figure}

    From~(\ref{geodKerr3}) one can obtain the permitted interval in~$r$ for
particles with $ \varepsilon = 1 $ and angular momentum $l = l_H - \delta $.
    To do this one must put the left hand side of~(\ref{geodKerr3})
to zero and find the root.
    In the second order in~$\delta$ close to the horizon one obtains
    \begin{equation} \label{KerrIntl}
l = l_H - \delta \ \ \Rightarrow \ \ \
x < x_\delta \approx x_H + \frac{\delta^2 x_C^2}{4 x_H \sqrt{1-A^2} } \,.
\end{equation}
    The effective potential for the case $ \varepsilon = 1 $
defined by the right hand side of~(\ref{geodKerr3})
    \begin{equation} \label{KerrVeff}
V_{\rm eff}(x,l) = - \frac{1}{2} \left( \frac{dr}{d \tau} \right)^2=
- \frac{1}{x} + \frac{l^2}{2x^2} - \frac{(A-l)^2}{x^3}
\end{equation}
(see, for example, Fig.~\ref{Veff})
%%%%%%%%%%%%%%%%%%%%%%%%%%%%%%%%%%%%%%%%%%%%%%%%%%
leads to the following behaviour of the particle.
    If the particle goes from infinity to the black hole it can achieve
the horizon if the inequality~(\ref{geodKerr5}) is valid.
    However the scattering energy in the centre of mass frame given
by~(\ref{KerrLimAMax}) is not large.
    But if the particle is going not from the infinity but
from some distance defined by~(\ref{KerrIntl}) then due to the form of
the potential it can have values of $l=l_H - \delta$
large than $l_R$ and fall on the horizon.
    If the particle falling from infinity with $ l \le l_R$ arrives
to the region defined by~(\ref{KerrIntl}) and here it interacts with other
particles of the accretion disc or it decays into a lighter particle which
gets an increased angular momentum $l_1 = l_H - \delta $,
then due to~(\ref{KerrLime1e2e}) the scattering energy in the centre of mass
system is
    \begin{equation} \label{KerrInEn}
E_{\rm c.m.} \approx \frac{1}{\sqrt{\delta}} \,
\sqrt{ \frac{ 2 m_1 m_2 (l_H - l_2) }{ 1- \sqrt{1 - A^2} }}
\end{equation}
    and it increases without limit for $\delta \to 0$.
    For $A_{\rm max} =0.998 $ and $l_2=l_L$, \
$ E_{\rm c.m.} \approx 3.85 m / \sqrt{\delta} $.  %% 3.85414

    Note that for rapidly rotating black holes $A= 1 - \epsilon$
the difference between $l_H$ and $l_R$ is not large
    \begin{eqnarray}
l_H - l_R &=& 2 \frac{\sqrt{1-A}}{A} \left( \sqrt{1-A} + \sqrt{1+A} -A \right)
\nonumber \\
&\approx& 2 (\sqrt{2}-1) \sqrt{\epsilon}\,, \ \ \ \epsilon \to 0 \,.
\label{KerrInDLR}
\end{eqnarray}
    For $A_{\rm max} =0.998 $, \ $ l_H - l_R \approx 0.04$   %% 0.0412465, 0.0370484
so the possibility of getting small additional angular momentum in
interaction close to the horizon seems much probable.
    The probability of multiple scattering in the accretion disc depends on
its particle density and is large for large density.

%\vspace{17pt}
%%%%%%%%%%%%%%%%%%%%%%%%%%%%%%%%%%%%%%%%%%%%%%%%%%%%%%%%%%%%%%%%%%%%%%
{\centering \section{\uppercase{\rm \large
Collision of particles inside Kerr black hole}}
\label{3secEnUFN}
}

    As one can see from formula~(\ref{KerrL1L2}) the infinite value
of the collision energy in the centre of mass system can be obtained inside
the horizon of the black hole on the Cauchy horizon~(\ref{HorCau}).
    Indeed, the zeroes of the denominator
in~(\ref{KerrL1L2}): $ x=x_H, \ x=x_C, \ x=0$.

    Let us find the expression for the collision energy for $x \to x_C$.
    Denote
    \begin{equation} \label{KerrlC}
l_C = \frac{2 \varepsilon x_C}{A} =
\frac{2 \varepsilon}{A} \left( 1 - \sqrt{1\!-\!A^2} \, \right)\!, \ \
l_{nC}=\frac{2 \varepsilon_n x_C}{A}.
\end{equation}
    Note that for $\varepsilon=1$
the Cauchy horizon can be crossed by the free falling from
the infinity particle under the same conditions on the angular
momentum~(\ref{geodKerr5}) as in case of the event horizon
and \ $ l_L < l_C \le l_R \le l_H$.

    To find the limit $r \to r_C$ for the black hole with a given angular
momentum~$A$ one must take in~(\ref{KerrL1L2}) $x = x_C + \alpha$
and do calculations with $\alpha \to 0 $ .
    The limiting energy has three different expressions depending on the values
of angular momenta.
    If
    \begin{equation} \label{lvnu1}
(l_1 - l_{1C}) (l_2 - l_{2C}) > 0 \,,
\end{equation}
    then
    \begin{eqnarray}
\frac{E_{\rm c.m.}^{\,2}(r \to r_C)}{2\, m_1 m_2} =
\frac{m_1^2 + m_2^2}{2 m_1 m_2} - \frac{l_{1C} l_{2C}}{4}
\nonumber \\
+\, \frac{1}{8} \left[ (l_{1C}^2 +4) \frac{l_{2C} - l_2}{l_{1C} - l_1} +
(l_{2C}^2 +4) \frac{l_{1C} - l_1}{l_{2C} - l_2}
\right],
\label{EcmVnu1}
\end{eqnarray}
    This formula is similar to~(\ref{KerrLimA})
if everywhere $H \leftrightarrow C$. \
    If
    \begin{equation} \label{lvnu00}
(l_1 - l_{1C}) (l_2 - l_{2C}) = 0 \,,
\end{equation}
for example, $l_1=l_{1C}$, then
    \begin{equation} \label{EcmVnu00}
\frac{E_{\rm c.m.}}{\sqrt{m_1 m_2}} \approx
\sqrt[4]{\frac{4 (l_2-l_{2C})^2 (\varepsilon_1^2 x_C + x_H)}{
x_C (x_H-x_C) (x-x_C)}}\,, \  \ x \to x_C.
\end{equation}
    If
    \begin{equation} \label{lvnu2}
(l_1 - l_{1C}) (l_2 - l_{2C}) < 0 \,,
\end{equation}
    then
    \begin{equation} \label{EcmVnu2}
\frac{E_{\rm c.m.}}{2 \sqrt{m_1 m_2}} \approx
\sqrt{\frac{ x_H (l_1 - l_{1C})(l_{2C} - l_2)}{
x_C (x_H - x_C) (x - x_C)}}, \  \ x \to x_C.
\end{equation}

    It seems that the limits of~(\ref{EcmVnu00}) and~(\ref{EcmVnu2})
are infinite for all values of angular
momenta $l_1,l_2$~(\ref{lvnu00}) and~(\ref{lvnu2}).
    This could be interpreted as instability of the internal Kerr's
solution~\cite{Lake10}.
    However, from Eq.~(\ref{geodKerr1}) we can see
    \begin{equation} \label{EcmVnuLimll}
\frac{d t}{d \tau}(x \to x_C + 0 ) = \left\{
\begin{array}{ll}
+ \infty, & \ {\rm if} \ \ l > l_C\,, \\
- \infty, & \ {\rm if} \ \ l < l_C\,.
\end{array} \right.
\end{equation}
    That is why the collisions with infinite energy can not be
realized (see also~\cite{Lake10}).

%%%%%%%%%%%%%%%%%%%%  my  %%%%%%%%%%%%%%%%%%%%%%%%%%%%%%%%%%%%%%%%%%%%
    For the particle falling to singularity in the equatorial plain
of the Kerr's black hole with $A \ne 0 $
    \begin{equation} \label{EcmSing}
\frac{d t}{d \tau}(x \to 0 ) = \left\{
\begin{array}{ll}
+ \infty, & \ {\rm if} \ \ l < \varepsilon A \,, \\
- \infty, & \ {\rm if} \ \ l > \varepsilon A \,.
\end{array} \right.
\end{equation}
    In case $ l = \varepsilon A , \, A \ne 0 $ the righthand side
of~(\ref{geodKerr3}) for massive particle is negative for $x \to 0$
and falling to singularity for such particle is impossible.
    So for particles colliding in the vicinity of singularity
one has $(l_1 - \varepsilon_1 A)(l_2 - \varepsilon_2 A )>0$.
    Then from~(\ref{KerrL1L2}) one gets
    \begin{eqnarray}
\frac{E_{\rm c.m.}(r \to 0)}{m_1 + m_2} = \hspace{40pt}
\nonumber \\
= \sqrt{ 1+ \frac{m_1 m_2}{(m_1+m_2)^2}
\frac{ (l_1 -l_2 + (\varepsilon_2 - \varepsilon_1) A )^2 }
{ (l_1 - \varepsilon_1 A )(l_2 - \varepsilon_2 A)}}.
\label{KerrLimSing}
\end{eqnarray}
    That is why collision of particles with infinite energy in the centre
of mass frame is impossible even in singularity.

\vspace{17pt}
%%%%%%%%%%%%%%%%%%%%%%%%%%%%%%%%%%%%%%%%%%%%%%%%%%%%%%%%%%%%%%%%%%%%%%

\end{document}